\documentclass[conference]{IEEEtran}
\usepackage{hhline}
\usepackage{url}
\usepackage{tabularx}
\usepackage{booktabs}
\usepackage{nameref}
\usepackage{hyperref}
\usepackage{listings} 
\usepackage{graphicx}
\usepackage{etoolbox} 
\usepackage{subcaption}
\usepackage{xargs,paralist,booktabs,array,tabularx,tikz,color,colortbl}
\usepackage{makecell}
\usepackage{rotating}
\usepackage{multirow}
\usepackage{amssymb}
\usepackage[colorinlistoftodos]{todonotes}
\definecolor{Gray}{gray}{0.9}
\definecolor{Lightgray}{gray}{0.96}
\DeclareTextFontCommand{\red}{\color{red}}

\usetikzlibrary{trees,shapes.geometric,shapes.symbols,chains,calc,positioning,arrows,decorations.pathmorphing,decorations.pathreplacing}

\newlength\bubblesize
\newcommand\yes{\tikz[baseline=0.1ex] \fill[black] (\bubblesize,\bubblesize) circle (\bubblesize);}

\newcommand\no{\tikz[baseline=0.1ex] \draw[black, line width=0.2ex] (\bubblesize,\bubblesize) circle (\bubblesize-.5\pgflinewidth);}

\newcommand{\ml}[2]{\multicolumn{#1}{l}{#2}}

\newcommand{\so}{Stack Overflow}
\newcommand{\googleplay}{Google Play}
\newcommand{\cp}{copy and paste}
\newcommand{\cpd}{copied and pasted}
\newcommand{\androidapps}{Android applications}

\begin{document}

\title{\huge\textbf{\so{} Considered Harmful?}\\ \LARGE{\textbf{The Impact of Copy\&Paste on Android Application Security}}}

\author{
\IEEEauthorblockN{\normalsize{Felix Fischer, Konstantin B{\"o}ttinger, Huang Xiao, Christian Stransky\IEEEauthorrefmark{1}, Yasemin Acar\IEEEauthorrefmark{1}, Michael Backes\IEEEauthorrefmark{1}, Sascha Fahl\IEEEauthorrefmark{1}}}
\IEEEauthorblockA{Fraunhofer Institute for Applied and Integrated Security; \IEEEauthorrefmark{1}CISPA, Saarland University}
}


\maketitle

\begin{abstract}

Online programming discussion platforms such as \so{} serve as a rich source of information for software developers. Available information include vibrant discussions and oftentimes ready-to-use code snippets. Previous research identified \so{} as one of the most important information sources developers rely on. Anecdotes report that software developers \cp{} code snippets from those information sources for convenience reasons. Such behavior results in a constant flow of community-provided code snippets into production software. To date, the impact of this behaviour on code security is unknown.

We answer this highly important question by quantifying the proliferation of security-related code snippets from \so{} 
in \androidapps{} available on \googleplay{}. 
Access to the rich source of information available on \so{} including ready-to-use code snippets provides huge benefits for software developers. However, when it comes to code security there are some caveats to bear in mind: Due to the complex nature of code security, it is very difficult to provide ready-to-use and secure solutions for every problem. 
Hence, integrating a security-related code snippet from \so{} into production software requires caution and expertise. Unsurprisingly, we observed 
insecure code snippets being copied into \androidapps{} millions of users install from \googleplay{} every day. 

To quantitatively evaluate the extent of this observation, we scanned \so{} for code snippets and evaluated their security score using a stochastic gradient descent classifier. In order to identify code reuse in \androidapps{}, we applied state-of-the-art static analysis. Our results are alarming: 15.4\% of the 1.3 million \androidapps{} we analyzed, contained security-related code snippets from \so{}. Out of these 97.9\% contain at least one insecure code snippet. 
\end{abstract}

\section{Introduction}

Discussion platforms for software developers have grown in popularity. Especially inexperienced programmers treasure the direct help from the community providing easy guide and most often even ready-to-use code snippets. It is widely believed that copying such code snippets into production software is generally practiced not only by the novice but by large parts of the developer community. Access to the rich source of information given by public discussion platforms provides quick solutions. This allows fast prototyping and an efficient workflow. Further, the public discussions by sometimes experienced developers potentially promote distribution of best-practices and may improve code quality on a large basis.

However, when it comes to code security, we often observe the opposite. Android-related discussions on \so{} for example include an impressive conglomeration of oddities: from requesting too many and unneeded permissions~\cite{Felt:2011kj} to implementing insecure X.509 certificate validation~\cite{Fahl:2012uh} to misusing Android's cryptographic API~\cite{Egele:2013:ESC:2508859.2516693}, a developer who is seeking help can find solutions 
for almost any problem. While such solutions oftentimes provide functional code snippets, many of them threaten code security. Those insecure code snippets commonly have a rather solid life-cycle: provided by the community, \cpd{} by the developer, shipped to the customer, and exploited by the attacker. To date it is unknown to what extent software developers \cp{} code snippets from information sources into production software. Is this phenomenon limited to just occasional instances, or is it rather a general and dangerous trend threatening code security to a large extent?

We answer this highly important question by measuring the frequency of 1,161 
insecure code snippets posted on \so{} that were copied and pasted into 1,305,820 
\androidapps{} available 
on \googleplay{}. We demonstrate that the proliferation of insecure code snippets within the Android ecosystem, and thus the impact of insecure code snippets posted on \so{}, poses a major and dangerous problem for Android application security.

\begin{figure*}[t]
	\centering
    \includegraphics[width=1\textwidth]{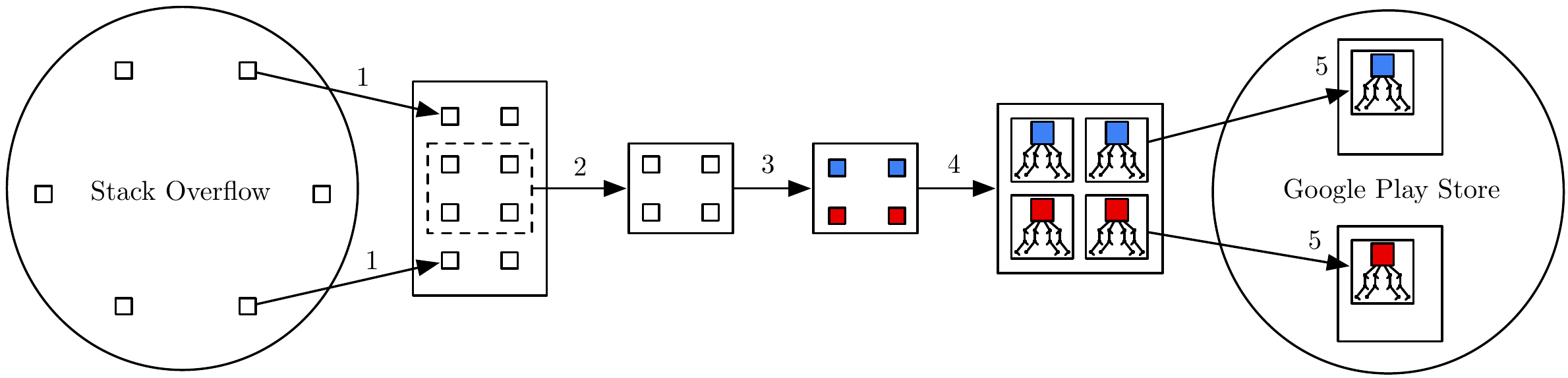}
	\caption{Overall processing pipeline of code extraction (1), filtering (2), classification (3), program dependency graph generation (4), and clone detection (5).}
	\label{fig:proc_pipe}
\end{figure*}

\subsection*{Our Contributions}
We investigate the extent 
security-related code snippets posted on \so{} were copied into \androidapps{} available 
on \googleplay{}. Our contributions can be summarized as follows:

\begin{itemize}
    \item We identified all Android posts on \so{}, extracted all (4,019)
    security-related code snippets and analyzed their security using a robust machine learning approach. 
    As a result we provide a security analysis for all security-related Android code snippets available on \so{}.
    \item We applied state-of-the-art static code analysis techniques to detect extracted code snippets from \so{} in 1.3 million \androidapps{}.
    \item We found that 15.4\% of all 1.3 million \androidapps{} contained security-related code snippets from \so{}. Out of these 97.9\% contain at least one insecure code snippet.
    \item We designed and implemented a fully automated large-scale processing pipeline for measuring the flow of security-related code snippets from \so{} into \androidapps{}.
    \item We make all data available on \url{https://www.aisec.fraunhofer.de/stackoverflow}.
\end{itemize}

Our processing pipeline is fully automated and designed to scale to extensive measurements of platforms other than \so{} and software repositories other than \googleplay{}. 

\section{Processing Pipeline Architecture}

In this section, we discuss the architecture of our processing pipeline. The individual steps of the processing pipeline are described in detail in subsequent sections.

As depicted in Figure \ref{fig:proc_pipe} the code originates in the \so{} database (on the left) and flows into \googleplay{} (on the right). To measure this flow we first crawl \so{} and extract every single code snippet in the database (1). From this comprehensive snippet collection we filter those that are security-related (2). We discuss steps (1) and (2) in detail in Section \ref{sec:parser} on code extraction and filtering. This provides us with a set of security-related snippets. In order to label each of them \textit{secure} or \textit{insecure} we define labeling rules as described in Section \ref{sec:scoring} and apply machine learning classification (3) using support vector machines (cf. Section \ref{sec:security_classification}). Next, we generate an abstract representation of each labeled code snippet (4) that allows us to detect their clones in \googleplay{} (5) (cf. Section \ref{sec:finding-code-snippets-in-apps}). Each step is fully automated and designed for large scale analysis. Only the training step for supervised machine learning classification (3) requires manual labeling of training data. However, this must be done only once for a small fraction of snippets, classification of very large sets of code snippets afterwards runs fully automated and is therefore just a matter of processing power and time. As we will show in the evaluation in Section \ref{sec:largescale} our proposed approach is time-efficient and yields decent results.

\section{Code Extraction and Filtering}
\label{sec:parser}
In the first step (1) of our processing pipeline we crawl discussion threads from a developer discussion platform for actual code snippets. In a second step (2) we extract the security-related fraction from the collected snippet set. We begin this section by defining the criteria for security-related code snippets and continue with describing both processing steps which allow us to extract security-related code snippets from \so{}.

\subsection{Security-related Code Snippets}
\label{sec:secrelcodsni}
On Android, security operations include but are not limited to cryptographic operations, secure network communication and transmission, validation via PKI-based mechanisms, as well as authentication and access control. For each of these operations, developers can select from a set of security APIs to perform them. We define code elements of these APIs as an indicator for security-related code. A code snippet is considered security-related if it makes calls to one of the following Java security libraries:~\cite{javase}
\begin{itemize}
    \item Cryptography: Java Cryptography Architecture (JCA), Java Cryptography Extension (JCE) 
    \item Secure network communications: Java Secure Socket Extension (JSSE), Java Generic Security Service (JGSS), Simple Authentication and Security Layer (SASL)
    \item Public key infrastructure: X.509 and Certificate Revocation Lists (CRL) in java.security.cert, Java certification path API, PKCS\#11, OCSP
    \item Authentication and access control: Java Authentication and Authorization Service (JAAS)
\end{itemize}

Additionally, we included code snippets with reference to the following security libraries, which were specially designed for Android: BouncyCastle (BC) is the default, pre-installed cryptographic service provider on Android and is widely used~\cite{Egele:2013:ESC:2508859.2516693}. 
SpongyCastle\footnote{cf. \url{https://rtyley.github.io/spongycastle/}} (SC) gives a repackaged version of BC which provides additional functionality. We looked for code snippets containing both BC and SC API calls.

Furthermore, we extracted code snippets for the Apache TLS/SSL package as part of the \texttt{HttpClient} library which is one of the most used libraries on GitHub~\cite{Zhitnitsky}.

We also included code snippets that reference security libraries specifically designed with usability in mind~\cite{5958047}, e.g. keyczar~\cite{dey2008keyczar} and jasypt~\cite{jasypt}, which were designed to simplify the safe use of cryptography for developers. 

To contrast Android's default providers and the usable security libraries with a more inconvenient alternative, we searched for snippets that use GNU Crypto. Although this library also implements a JCA provider, it is difficult to integrate into Android~\cite{González2015}, which makes it interesting to see how it is being discussed on \so{}, as well as whether or not developers use it.

Table \ref{tab:sec-libs} lists the considered security libraries and gives an overview of their supported features.

\begin{table}
  \centering
  \footnotesize
  \setlength{\tabcolsep}{2pt}
  \def\arraystretch{1.2}
  \begin{tabular}{@{}p{2.6cm} *{9}{c >{\columncolor{Gray}}c}}\\
    \textbf{} &
    \textbf{} &
    \tikz \node[rotate=90] {TLS}; &
    \tikz \node[rotate=90] {Symmetric Cryptography}; &
    \tikz \node[rotate=90] {Asymmetric Cryptography}; &
    \tikz \node[rotate=90] {Secure Random Number Generation}; &
    \tikz \node[rotate=90] {Message Digests}; &
    \tikz \node[rotate=90] {Digital Signatures}; &
    \tikz \node[rotate=90] {Authentication}; &
    \tikz \node[rotate=90] {Usability by Design}; 
    \\
    \hline
    \textbf{Standard API}   & &\yes{} &\yes{} &\yes{} &\yes{} &\yes{} & \yes{} & \yes{} & \no{}\\
    \textbf{BouncyCastle}   & &\yes{} &\yes{} &\yes{} &\yes{} &\yes{} & \yes{} & \yes{} & \no{}\\
    \textbf{SpongyCastle}   & &\yes{} &\yes{} &\yes{} &\yes{} &\yes{} & \yes{} & \yes{} & \no{}\\
    \textbf{Apache TLS/SSL} & &\yes{} &\no{}  &\no{}  &\no{}  &\no{}  & \no{}  & \no{}  & \no{}\\
    \textbf{keyczar}        & &\no{}  &\yes{} &\yes{} &\no{} &\no{} & \yes{} & \yes{} &\yes{}\\
    \textbf{jasypt}         & &\no{}  &\yes{} &\no{}  &\no{}  &\yes{} & \no{} & \no  &\yes{}\\
    \textbf{GNU Crypto}     & &\no{}  &\yes{} &\yes{} &\yes{} &\yes{} & \yes{} & \no{} & \no{}\\
    \hline
    
    \multicolumn{10}{c}{\makecell{\yes{} = fully applies; \\\no{} = does not apply at all}} \\
    \multicolumn{10}{c}{} \\
  \end{tabular}
  \caption{Cryptographic libraries and their supported features.}
  \label{tab:sec-libs}
\end{table}

\subsection{Finding Security-related Code Snippets on \so{}}
\label{sub:security-related-snippets}
Code snippets on \so{} are surrounded by \texttt{<code>} tags and can therefore easily be separated from accompanying text and extracted.

We were interested in security-related code snippets only. Therefore, we filtered code snippets that contain API elements of the security libraries described in Section~\ref{sec:secrelcodsni}. In order to decide to which API a code element belongs, we need Fully Qualified Names (FQN) (i.e. package names in Java) of elements in code snippets. Since class and method names are not unique, different API can contain classes (e.g. \texttt{android.util.Base64}, \texttt{java.util.Base64}) and methods (e.g. \texttt{java.security.Cipher.getInstance}, \texttt{java.security.Signature.getInstance}) which share the Partially Qualified Name (PQN). Therefore, it is necessary to know the unique FQN to be able to disambiguate code elements. 

Code snippets posted on \so{} are often incomplete or erroneous and therefore only PQNs may be available. Since disambiguating partial Java programs is an undecidable problem~\cite{Dagenais:2008:ESA:1449764.1449790}, we used an oracle called JavaBaker~\cite{Subramanian:2014:LAD:2568225.2568313} to decide to which API a code element belongs. The oracle consists of a user-defined set of APIs which is used to apply a constraint-based approach to disambiguate types of given code elements. Given a code snippet JavaBaker returns the FQN for each element in the code, if it belongs to one of the initially given libraries. The JavaBaker oracle has a precision of 0.97 and a recall of 0.83~\cite{Subramanian:2014:LAD:2568225.2568313}. It is not restricted to specific libraries which allowed us to apply it for our use case. With JavaBaker, using the security libraries explained in Section \ref{sec:secrelcodsni} as the user-defined set of APIs, we were able to determine to which of the given security APIs a type reference, method call, or field access in a code snippet belongs. A code snippet is therefore considered security-related if the returned result of the oracle is not empty. We apply this to filter security-related code snippets from \so{}.

Since the security APIs might contain packages whose usage does not indicate implementation of security code (e.\,g. \textit{util} or \textit{math} packages), our snippet filter includes a blacklist to ignore those non-security-related packages. We compiled this blacklist manually by inspecting each package individually. 

Code snippets may contain sparsely used code elements. For instance, an object can be declared and initialized, but not used subsequently in the snippet. In this case, the oracle only has the PQN of the element and the call to the constructor as information to decide the FQN. This can lead to false positives because the oracle has insufficient information to narrow down possible candidates.
To give an example, the oracle reported \texttt{java.security.auth.login.Configuration} as the FQN for a code element with type \texttt{Configuration} whose true FQN was \texttt{android.content.res.Con- figuration}.
The related object only made a call to the constructor, hence it was impossible to disambiguate the given type \texttt{Configuration}. Luckily, these false positives are easily detectable by filtering out snippets for which the oracle reports the \texttt{<init>} method only or no methods at all. We do not worry about true positives we might sort out this way, as we are not interested in code snippets that contain security elements which are not used after initialization.

\subsection{Limitations}
\label{sec:rationale-oracle}
The main purpose of the the oracle-based filter is to decide whether a given snippet is security-related. As it does this by examining the snippet for utilization of the defined security libraries, it might label a snippet as security-related, even though it does not belong to a security context. This is the case if an API element which is heavily used for security purposes can also be used in a non-security context. For instance, in a security context snippets would use hashing algorithms for verifying data integrity. In a non-security context hashes may be used for data management purposes only. In both cases the snippet would reference elements of one of the given security APIs which causes the filter to label the snippets as security-related.

\section{Code labeling}
\label{sec:scoring}
Now that we have extracted security-related code snippets (cf. Figure \ref{fig:proc_pipe}, (1) and (2)), we need to classify them as such. Therefore, we first provide the label definition and labeling rules and give details on the actual machine learning based classification in Section \ref{sec:security_classification}. We apply supervised learning and therefore need to manually label a small fraction of extracted code snippets to train the support vector machine. Therefore a pair of two reviewers inspected the set of 1,360 security-related snippets extracted from answer posts from \so{}. In case of conflicts, a third reviewer was consulted and the conflict was resolved (by explaining the reasoning of the reviews).

To better understand which topics were discussed (in combination with code snippets) on \so{}, we categorized each code snippet into one or multiple of the following categories: \textit{SSL/TLS}, \textit{Symmetric cryptography}, \textit{Asymmetric cryptography}, \textit{One way hash functions}, \textit{(Secure) Random number generation}. 

\subsection{Security Labels}
We checked whether or not code snippets were security risks when pasted into Android application code and labeled them either secure or insecure:

\begin{itemize}
    \item[\textbf{Secure}]~\par
\end{itemize}
\begin{itemize}
        \item Snippets that contain up-to-date and strong algorithms for symmetric cryptography~\cite{sheffer2015p,rfc2898}, sufficiently large keys for RSA or elliptic curve cryptography~\cite{Manger:2001:CCA:646766.704143, nist-recommend} or secure random number generation~\cite{Egele:2013:ESC:2508859.2516693}. 
        \item Snippets that contain code that does not adhere to security best practices, but does not result in easily exploitable vulnerabilities either, e.g. usage of RSA with no or PKCS1 padding~\cite{Bleichenbacher:1998:CCA:646763.706320}, SHA1 
        or outdated versions of SSL/TLS~\cite{sheffer2015p}. 
        \item Snippets that contain code whose security depended on additional developer input, e.g. the symmetric cryptography algorithm or key size is a parameter, which is configurable by the developer.
\end{itemize}
\begin{itemize}
    \item[\textbf{Insecure}]~\par
\end{itemize}
\begin{itemize}
        \item Snippets that contained obviously insecure code, e.g. using outdated algorithms or static initialization vectors and keys for symmetric cryptography, weak RSA keys for asymmetric cryptography, insecure random number generation~\cite{Egele:2013:ESC:2508859.2516693}, or insecure SSL/TLS implementations~\cite{Fahl:2012uh}.
\end{itemize}
This labeling is very conservative as it classifies only the definitely vulnerable code snippets as insecure.

\subsection{Labeling Rules}
\label{sec:labeling-rules}
Code security was investigated for the category specific parameters, which are introduced in this section. Based on these parameters we state a security metric which provides the rules for labeling the code snippets. Our stated security metric does not intend to be an exhaustive metric for each security category, but only considers security parameters which were actually used in the snippets of our corpus. In the following, we provide tables for each category which depict secure and insecure parameters for quick lookup. Additionally, we give details on parameters that were ambiguous or need further explanation. We defined the following labeling rules for security classification:

\subsubsection{SSL/TLS}
\begin{table}[htb!]
\centering
\begin{tabular}{lcc} \hline \hline
Parameter   &   Secure  & Insecure                                                                       \\ \hline \hline
Hostname   &    browser compatible,                     &   allow all                                   \\ 
Verifier   &    strict                                  &   hosts~\cite{fahlccs:2013}               \\ \hline
Trust      &    default,                                &   trust all~\cite{Fahl:2012uh},                                   \\    
Manager    &    secure                                  &   bad pinning~\cite{Oltrogge:usenix15,fahlccs:2013},          \\
           &    pinning                                 &   validity only                                   \\ \hline
Version    &    $>=$TLSv1.1~\cite{sheffer2015p}            &   $<$TLSv1.1~\cite{turner2011prohibiting, sheffer2015p,dierkstransporttls13,dierkstransporttls12}           \\  \hline
Cipher     &    DHE\_RSA, ECDHE                         &   RC4,3DES,           \\
Suite       &   AES$>=$128, GCM                           &   AES-CBC             \\
            &   SHA$>=$256~\cite{sheffer2015p}            &   MD5, MD2~\cite{sheffer2015p, Vaudenay:2002:SFI:647087.715705}            \\ \hline
OnReceived- &   cancel                                   &   proceed             \\
SSLError    &                       &                       \\ \hline \hline
\end{tabular}
\caption{Secure and insecure TLS parameters.}
\label{table:ssl-metric}
\end{table}
Table~\ref{table:ssl-metric} illustrates the TLS parameters we investigated~\cite{Fahl:2012uh}. The \texttt{HostnameVerifier} checks whether a given certificate's common name matches the server's hostname. \texttt{TrustManager} implementations allow developers to implement custom certificate (chain) validation strategies. Insecure hostname verifier or trust manager implementations make an application vulnerable to Man-In-The-Middle attacks. According to~\cite{Fahl:2012uh} we labeled \texttt{TrustManager} and \texttt{HostnameVerifier} implementing insecure validation strategies as insecure. \texttt{TrustManagers} that implement public key or certificate pinning are considered secure. However, we label pinning as insecure if the pinset contains ambiguous values, e.\,g. serial number of the certificate~\cite{Oltrogge:usenix15,fahlccs:2013}. We also investigated TLS security of \texttt{WebViews}. Developers can implement their own \texttt{OnReceivedSSLError} method to handle certificate validation errors while loading content via HTTPS and can ignore validation errors by proceeding the TLS handshake. \hfill\\

\subsubsection{Symmetric Cryptography}
\label{par:symcrypto}
\begin{table}[htb!]
\centering
\begin{tabular}{lcc} \hline \hline
Parameter                   &   Secure                  & Insecure                                                                                      \\ \hline \hline
Cipher/Mode                 &   AES/GCM~\cite{sheffer2015p}                 & RC2~\cite{Kelsey:1997:RCB:646277.687180}, RC4~\cite{alfardan2013security},                    \\
                            &   AES/CFB~\cite{sheffer2015p}                 & DES~\cite{Kelsey:1997:RCB:646277.687180}, 3DES~\cite{Lucks1998},                        \\
                            &   AES/CBC*                & AES/ECB~\cite{Egele:2013:ESC:2508859.2516693},          \\
                            &                           & AES/CBC**~\cite{Vaudenay:2002:SFI:647087.715705}          \\
                            &                           & Blowfish~\cite{vaudenay1996weak, kara2007new}                                                 \\ \hline 
Key                         &   provider                & static~\cite{Egele:2013:ESC:2508859.2516693},       \\     
                            &   generated               & bad derivation~\cite{Egele:2013:ESC:2508859.2516693} \\ \hline
Initialization Vector       &   provider                & zeroed~\cite{Egele:2013:ESC:2508859.2516693},       \\   
(IV)                        &   generated               & static~\cite{Egele:2013:ESC:2508859.2516693},       \\
                            &                           & bad derivation~\cite{Egele:2013:ESC:2508859.2516693},\\ \hline
Password Based              & $>=$1k iterations~\cite{rfc2898},         & $<$1k iterations~\cite{rfc2898},\\ 
Encryption (PBE)            & $>=$64-bit salt~\cite{rfc2898},           & $<$64-bit salt~\cite{rfc2898}     \\
                            & non-static salt~\cite{rfc2898}           & static salt~\cite{Egele:2013:ESC:2508859.2516693}       \\ \hline \hline                   
\end{tabular}
\caption{Secure and insecure symmetric cryptography parameters.}
\label{table:sym-metric}
\end{table}
We investigated snippets for symmetric cryptography parameters as illustrated in Table~\ref{table:sym-metric}. We labeled \texttt{Ciphers} and \texttt{Modes} of operation which are known security best practices as secure. Ciphers and modes with known practical attacks were labeled insecure. The AES encryption mode CBC is depicted in both columns secure and insecure because known padding oracle attacks are only feasible in a client/server environment. If this encryption mode is used in a different scenario, we consider it as secure~\cite{Vaudenay:2002:SFI:647087.715705}. 
We labeled cryptographic \texttt{Keys} and \texttt{IV} which were statically assigned, zeroed or directly derived from text (such as passwords) as insecure~\cite{Egele:2013:ESC:2508859.2516693}.

\subsubsection{Asymmetric Cryptography}
\begin{table}[htb!]
\centering
\begin{tabular}{lcc} \hline \hline
Parameter                   &   Secure                                      &   Insecure      \\ \hline \hline
Cipher/Mode                 &   RSA                                         &                   \\
                            &   RSA/ECB                                     &                   \\
                            &   RSA/None                                    &                   \\ \hline
Padding                     &   PKCS1*,                                     &   PKCS1**         \\
                            &   PKCS8,                                      &                   \\
                            &   OAEPWithSHA-256                             &                   \\
                            &   AndMGF1Padding,                             &                   \\ \hline
Key                         &   RSA $>=$ 2048 bit                             &   RSA $<$ 2048 bit~\cite{Manger:2001:CCA:646766.704143}      \\
                            &   ECC $>=$ 224 bit                              &   ECC $<$ 224 bit~\cite{nist-recommend}                               \\ \hline \hline
\end{tabular}
\caption{Secure and insecure asymmetric cryptography parameters.}
\label{table:asym-metric}
\end{table}
We investigated snippets for asymmetric cryptography parameters as illustrated in table \ref{table:asym-metric}. The JCE API provides different \texttt{Cipher} and \texttt{Mode} transformation strings for RSA which include the definition of a block mode, e.\,g. RSA/ECB. However, these modes are ignored by the underlying provider and have no implication on security~\cite{javasecnadreltop}.
For RSA, we consider the used \texttt{Padding} and \texttt{Key} length to evaluate security~\cite{Manger:2001:CCA:646766.704143}.
We distinguish between a client/server and a non-client/server scenario. Only in the first scenario PKCS1 padding is vulnerable to padding oracle attacks and seen as a secure padding otherwise~\cite{Bleichenbacher:1998:CCA:646763.706320}. Secure and insecure key lengths for RSA and Eliptic curve cryptography~\cite{Manger:2001:CCA:646766.704143, nist-recommend} are shown in table \ref{table:asym-metric}.

\subsubsection{One Way Hash Functions}
\begin{table}[htb!]
\centering
\begin{tabular}{lcc} \hline \hline
Parameter                   &   Secure                                      &   Insecure                    \\ \hline \hline
PBKDF                       &   [PBKDF2](Hmac)                              &   [PBKDF2](Hmac)              \\
                            &   $>=$SHA224~\cite{keylength}                   &   MD2, MD5~\cite{keylength}   \\  \hline 
Digital Signature           &   $>$SHA1                                       &   MD2, MD5                    \\ \hline
Credentials                 &   $>$SHA1                                       &   MD2, MD5                    \\ \hline \hline
\end{tabular}
\caption{Secure and insecure hash function parameters}
\label{table:owhf-metric}
\end{table}
We investigated snippets for one way hash function parameters, as illustrated in Table~\ref{table:owhf-metric}, in the context of password-based key derivation, digital signatures, and authentication/authorization. These were the only categories where code snippets from our analysis corpus made explicit use of hash functions. 
In the context of OAuth and SASL (authentication and authorization), attacks are mainly possible through flaws in website implementations~\cite{SANSInstitute}. Therefore, we only analyzed which hashing schemes were used for hashing credentials.

\subsubsection{(Secure) Random Number Generation}
\begin{table}[htb!]
\centering
\begin{tabular}{lcc} \hline \hline
Parameter                   &   Secure                                      &   Insecure                \\ \hline \hline
Type                        &   SecureRandom                                &   Random                  \\  \hline
Seeding                     &   nextBytes,                                  &   setSeed-$>$nextBytes,      \\ 
                            &   nextBytes-$>$setSeed                        &   setSeed with            \\
                            &                                               &   static values~\cite{Egele:2013:ESC:2508859.2516693}           \\ \hline \hline
\end{tabular}
\caption{Secure and insecure parameters for (secure) random number generation.}
\label{table:sec-rand-metric}
\end{table}
We investigated snippets for (secure) random number generation parameters shown in table \ref{table:sec-rand-metric}. The main problem which can lead to security problems lies in provider specific implementation and ambiguous documentation of manual seeds~\cite{java-seeding}.
We conclude that besides calling \texttt{nextBytes} only, which lets \texttt{SecureRandom} seed itself, calling \texttt{nextBytes} followed by \texttt{setSeed} is a secure sequence because \texttt{SecureRandom} is still self-seeded. The latter call to \texttt{setSeed} just supplements the seed and does not replace it~\cite{java-seeding}. Without calling \texttt{nextBytes} first, a call to \texttt{setSeed} may completely replace the seed. This behavior differs between several providers and is often ill-described in official documentation~\cite{java-seeding}. Therefore, we consider this call sequence as insecure if an insufficient seed is given.

\subsection{Limitations}
\label{sec:limit_2}
Our code snippet reviews might be limited in multiple ways in this step. Although we based our review decisions on widely accepted best practices and previous research results and let multiple reviewers review all snippets we cannot entirely eliminate incorrect labeling. The security of most code snippets depends on input values (e.\,g. initialization parameters) that were not given in all code snippets. Therefore, our results might under- or overreport the prevalence of insecure APIs in Android applications.

\section{Code Classification}
\label{sec:security_classification}
In this section, we present our method for large-scale code snippet classification, which corresponds to (3) in the overall processing pipeline (cf. Figure \ref{fig:proc_pipe}). 

Manual snippet analysis allows profound insight into security problems specifically raised from crowd-sourced code snippets. Further, it allows the creation of a rich data set that annotates crowd-sourced code snippets from \so{}. This opens the doors for machine learning based classification. To the best of our knowledge, we are the first to contribute such a data set to the machine learning community. 

The security scoring of code snippets can be seen as a classification problem, which we can effectively solve by a variety of classifiers, e.\,g. feed-forward neural networks, decision trees, support vector machines, and many more. By manually labeling a subset of the collected snippets as secure and insecure (cf. Section~\ref{sec:scoring}), we are able to produce a training data set for binary classifiers. The trained model is then applied to classify unknown code snippets. We apply the binary classifier on all security-related snippets extracted by the oracle-based filter to provide an automatic procedure of security assessment. 

Note that it is arguable that machine learning based methods would deliver more benefits than rule-based methods on solving security problems. 
Our binary classifier can efficiently extract discriminative information from the data set we collected, which might be overlooked by rule based methods. 
The features rely merely on the vocabulary level of input code snippets, without even understanding how they are functioning. 
This allows the model to easily scale up to thousands of snippets in a few seconds which is not affordable for manual processing.
\begin{figure}[h!]
	\centering
	\includegraphics[width=.35\textwidth]{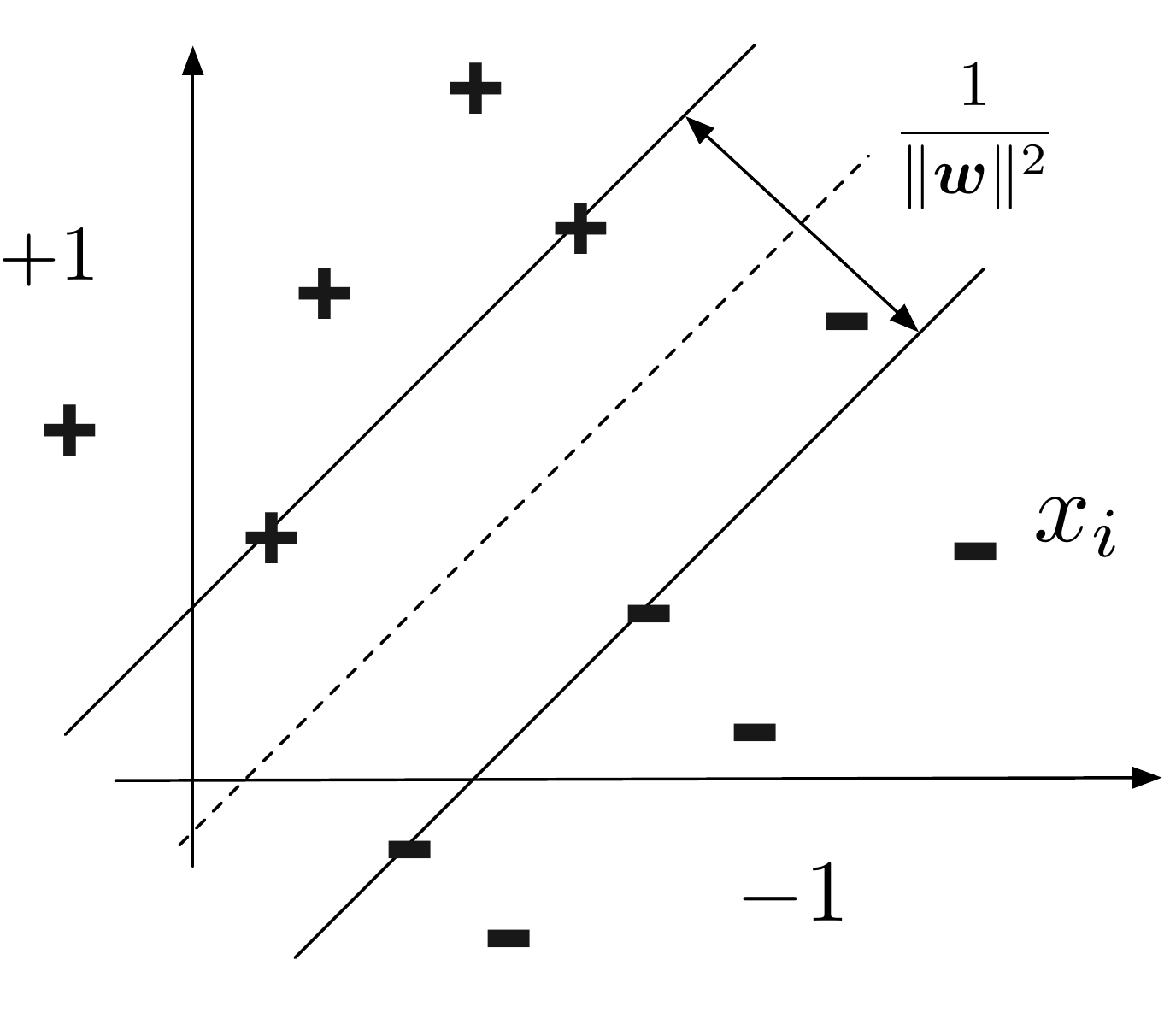}
	\caption{Illustration of SVM binary classifier. It maximizes a margin $ \frac{1}{\|\mathbf{w}\|^2}$ to separate positive and negative samples in its correct side. Note that a small portion of data samples are allowed within the margin, which can be controlled by a set of slack variables $ \mathbf{\xi} $.}
	\label{fig:svm_example}
\end{figure}

\subsection{Support Vector Machine}
We employ the binary classifier Support Vector Machine (SVM) as our learning model. 
In our scenario, the labeled training data set contains two classes, namely, insecure and secure code snippets. The collected code snippets can be regarded as documents. We argue that discriminative patterns can be discovered by examining the tokens in code snippets. These can be any combination of alphabets and symbols, e.g., \textit{while}, \textit{return}. Therefore, in our setting the learning problem is a document classification problem with binary classes from a set of tokens.

Given a training dataset of $n$ samples $\mathcal{X}=\left\lbrace x_i\right\rbrace_{i=1}^n$, and its corresponding labels $\left\lbrace y_i\right\rbrace_{i=1}^n \in [+1, -1]$, a SVM classifier learns a margin that maximally separates training samples into two classes as illustrated in Figure \ref{fig:svm_example}. The objective function can be formulated as follows,
\begin{eqnarray}
& \min_{w,b,\xi} \frac{1}{2}w^Tw + C\sum_{i=1}^n \xi_i \label{eq:svm_obj} \\
\mbox{s.t.} & y_i (w^T \phi(x_i) + b) \geq 1-\xi_i \nonumber \\
& \xi_i \geq 0, i=1,\ldots, n \nonumber 
\end{eqnarray}
In (\ref{eq:svm_obj}) we note that minimizing $w$ equals maximizing a margin. 
SVM introduces a set of slack variables $\left\lbrace \xi_i \right\rbrace$ to soften the margin, such that a small portion of training samples are allowed to be misclassified. Importantly, we also note that the feature mapping $\phi(x_i)$ defined over $\mathcal{X}$ can intrinsically handle non-linear cases by the so called 'kernel trick'. For more details, we refer to~\cite{Scholkopf00b}.

\subsection{Feature Extraction}
\label{sec:tfidf}
Since the learning problem of detecting the security level of code snippets can be viewed as a document classification problem, we employ a common feature extraction method named \textit{tf-idf} vectorizer \cite{Wu:2008:ITT:1361684.1361686}. The \textit{tf-idf} vectorizer transforms the whole set of code snippets into a numeric matrix. Each of the code snippets is considered as a document, namely an input data sample. We compute term frequency (\textit{tf}) and inverse document frequency (\textit{idf}) with respect to the total number of snippets. 

For each snippet, the term frequency is computed by counting each token 
within its document. For the inverse document frequency, we compute the inverse of the number of documents where each token appears in. Then the \textit{tf-idf} score is simply a multiplication of term frequency and inverse document frequency. In the end, we maintain a vocabulary of code tokens parsed from the snippets. This vocabulary will be converted into a numeric vector of a fixed length containing all possible tokens' frequency in this snippet. 
Normally, \textit{tf-idf} vectorizer will form a high dimensional sparse data set with many entries being set to zero, if all the individual tokens are taken into account. Some tokens, e.g., randomly generated numbers, variable and class names, only appear in particular documents and therefore their document frequency is quite low. Document frequency can be very high for other tokens, e.g., common language terms such as \textit{return}, \textit{public}.
The \textit{tf-idf} scores for these tokens will be automatically justified by the inverse document frequency,
such that their contribution to the discriminative function will also be reweighed. Finally, the sparse data set is then fed to SVM as the training data set. We expect the tokens found in each snippet to represent an encoding of how secure the code snippet will be.

\section{PDG Generation and Code Detection}
\label{sec:finding-code-snippets-in-apps}
Our processing pipeline has now filtered security-related code snippets from \so{} and classified them either as secure or insecure (cf. Figure \ref{fig:proc_pipe}, (1) to (3)). Next, we aim to detect these code snippets in compiled Android applications from \googleplay{}, (cf. Figure \ref{fig:proc_pipe}, (4) and (5)). 

Snippets are given as source code and \androidapps{} are only available as high-level binaries (i.\,e. DEX files). To be able to apply static code analysis techniques, code snippets and Android applications must be transformed into the same (intermediate) representation (IR). In this section we first describe this transformation step (4) and then give a detailed explanation of the method we apply to detect code snippets in Android applications (5).

\subsection{Code Snippet Compiling}
\label{sec:}
Static code analysis tools require complete programs to work properly~\cite{Dagenais:2008:ESA:1449764.1449790}. Most code snippets from \so{} however are not complete programs. They mostly do not compile without error since required method or class information is missing~\cite{Subramanian:2014:LAD:2568225.2568313}. A snippet may be a subset of a larger program which is not accessible or additional dependencies (e.\,g. external libraries) might me unknown~\cite{Dagenais:2008:ESA:1449764.1449790}. 

For incomplete code snippets creating a typed and complete IR is difficult. To overcome this, we use Partial Program Analysis (PPA)~\cite{Dagenais:2008:ESA:1449764.1449790}. It was specifically designed to create complete and typed abstract syntax trees (AST) from source code of partial Java programs.
PPA is able to resolve syntactic ambiguities which often times arise in code snippets. For example, the statement \texttt{SSLSocketFactory.getDefault()} does not allow to decide if \texttt{SSLSocketFactory} is a class or field name if not explicitly declared. In this specific case, \texttt{SSLSocketFactory} is a missing class and therefore \texttt{getDefault()} should be resolved to a static method call.
PPA is also able to disambiguate possible typing problems which arise in case not all declared types are available. This is done by reconstructing data types from the snippets without having access to source files, binaries or libraries.
For data types that cannot be resolved applying PPA, a generic data type \texttt{UNKNOWNP.UNKNOWN} is used. This ensures that the created AST remains complete.

To transform snippets and applications into the same IR we use WALA\footnote{http://wala.sourceforge.net}. Since WALA operates on JVM bytecode, we transform Android applications to JVM bytecode using enjarify~\cite{enjarify} first. To be able to operate on \so{} code snippets, we modified WALA by integrating PPA. This allows us to transform incomplete code snippets into WALAs IR.
Before transformation, we make sure code snippets represent a complete Java class. Based on these snippets, we create the complete and typed AST using PPA. We were able to successfully process 1,293 answer (85.2\%) and 1,668 question snippets (66.6\%). Snippets which could not be compiled mostly had a too erroneous syntax and were therefore rejected by WALA. Furthermore, a lot of snippets contained a mixture of Java code and non-commented text (e.\,g code blocks were replaced with '(...)'). We ignored those snippets for further analyses~\cite{Subramanian:2014:LAD:2568225.2568313}.

\subsection{Code Snippets in Apps}
\label{sub:reused-code-snippets}
Code snippet containment is given if an application contains code that is very similar to the code snippet. However, a full match is not necessary. Instead we use a detection algorithm which is robust to \textit{fractional} and \textit{non-malicious} modifications\footnote{Code obfuscation is not intended to be covered by our approach}. 

We base code snippet detection on finding similar Program Dependency Graphs (PDG) which store data dependencies by applying a modified approach of Crussel et al.~\cite{Crussell2012}. 
They create PDGs for each method and define the independent subgraphs of a PDG as the basic code features that are considered for reuse detection. A method's PDG may contain several data independent subgraphs which are called semantic blocks. Code similarity is defined on the amount of similar semantic blocks that are shared among the compared code. 
Following this approach provides robustness to high-level modifications and trivial control-flow alterations, as well as non-malicious code insertions/deletion, code reordering, constants modifications and method restructurings as described in~\cite{Crussell2012}.

Though this approach allows the detection of reused code that has been subject to the defined modifications, we consider some of the given robustness features as inappropriate for our use case. It has several drawbacks when applied on detecting reuse of code snippets in relatively large applications. Many snippets are quite small in terms of lines of code and therefore result in small PDGs. In this scenario, different code might result in identical PDGs by chance. Therefore, we apply a more strict approach which additionally compares constants and method names that belong to a semantic block. This is reasonable because constants are critical for initializing Android security APIs. For instance cryptographic ciphers or TLS sockets are selected by using a transformation String (e.\,g. AES, TLS). Critical information like cryptographic keys, key lengths, initialization vectors, passwords and salts can be statically assigned in the code. 

To be able to detect reused constants they must not have been modified. Additionally, we compare method names that are part of a semantic block and belong to APIs of our predefined set of security libraries. This allows to distinguish security-related parts of the code, in case of different code with identical semantic blocks and empty or identical constant sets. Finally, we disallow class and method restructuring. This is necessary because we have have to ensure that detected semantic blocks are contained in classes and methods that have the same structure as the snippet. We compare semantic blocks, constants and method names on a per method base and ensure (nested) class membership by analyzing path names of all detected methods.



To avoid computational overhead, we limit the number of classes to search for code snippets to classes that contain security-related API calls as defined in Section~\ref{sub:security-related-snippets}.

Finding subgraph isomorphisms in PDGs is NP-hard and therefore not applicable for large-scale analysis~\cite{Crussell2013}. Therefore, we follow the approach of embedding graphs in vector spaces in order to reduce the problem of finding similar graphs to the problem of finding similar vectors~\cite{doi:10.1142/S021800140900748X}. We apply the embedding algorithm provided by Crussel et al.~\cite{Crussell2013} which assigns a semantic vector to each semantic block. The semantic vector stores information about nodes and edges, i.\,e. the overall structure of a semantic block. 
Nodes represent instructions, edges represent data dependencies between instructions. 

Each instruction type as provided by WALAs IR has two corresponding fields in the vector. One field stores node and the other stores edge information. The count of nodes for each instruction type (e.\,g. \textit{invokevirtual}, \textit{getfield}, \textit{new} or \textit{return}) in the semantic block is stored in the related nodes field of the instruction type in the vector. The maximum out node degree for each instruction type is used to store information about PDG edges. It holds the maximum count of outgoing edges over all nodes in a semantic block for a given instruction type and is stored in the related edges field of an instruction type in the vector.

To decide if two semantic vectors are similar, we calculate their Jaccard similarity~\cite{Hanna:2012:JSS:2481803.2481809, Jiang:2007:DSA:1248820.1248843} which describes the similarity ratio of two sets. Jaccard similarity for sets represented as binary vectors $X,Y$ is defined as $J_{s}(X,Y)=\frac{\sum_{i}(X_{i}\wedge Y_{i})}{\sum_{i}(X_{i}\vee Y_{i})}$. However, since the semantic vector stores count information of nodes and edges belonging to a semantic block, we define Jaccard similarity as $J_{s}(X,Y)=\frac{\sum_{i}min(X_{i},Y_{i})}{\sum_{i}max(X_{i},Y_{i})}$. Hence, two statements of the same instruction type in the semantic block represent different elements in the set representation of the semantic block. This is also true for outgoing edges which belong to the maximum out node degree. Therefore, two outgoing edges of a single node are different elements in the set representation. Furthermore, this definition ensures that only elements of the same instruction type are compared.

PPA is able to create an IR of an incomplete code snippet with an average correctness of 91\%~\cite{Dagenais:2008:ESA:1449764.1449790}. This gives us a threshold for Jaccard similarity of 0.91.
To decide if method names and constants of a semantic block are contained in another semantic block, we calculate their Jaccard containment. Jaccard containment depicts the containment ratio of an arbitrary set $X$ in another set $Y$ and is defined as $J_{c}(X,Y)=\frac{\left\vert{X\land Y}\right\vert}{\left\vert{X}\right\vert}$. We calculate both Jaccard containment of two method name and constant sets to evaluate whether all methods or constants are contained. We rely on a Jaccard containment value of 1.0 to satisfy the requirements of \ref{sub:reused-code-snippets}. We define containment of a code snippet in an app iff the following holds for each method in the snippet: 
\begin{itemize}
    \item For all given semantic blocks we find semantic blocks that satisfy Jaccard similarity and are contained in a single method contained in the callgraph of a given application.
    \item The method name set is fully contained in the same method.
    \item The constants set is fully contained in the same method.
    \item They belong to the same (nested) class.
\end{itemize}

\subsection{Exotic Case}

Empty TrustManager implementations require special treatment. They exclusively consist of overwritten methods (e.\,g. cf. Listing~\ref{listing-empty-tm}). These methods are mostly empty which means their PDG and methods and constants sets are also empty. Therefore, our approach cannot distinguish these methods. To avoid false positives, the TrustManager's methods \texttt{checkClientTrusted}, \texttt{checkServerTrusted} and \texttt{getAcceptedIssuers} receive special treatment. In case an empty method has been detected in the call graph of an application, we compare the method's fully qualified name with the method names given above. This way, we can successfully identify empty TrustManager implementations without false positives.


\section{Evaluation}
\label{sec:evaluation}
In this section we present a detailed evaluation of our approach. We discuss benchmarks and numbers for each step of our processing pipeline (cf. Figure \ref{fig:proc_pipe}). Further, we compare our results with feedback from the \so{} community, provided in the respective code threads of copied insecure snippets.

\subsection{Evaluation of Code Extraction and Filtering}
\label{sec:eval_code_extract_filter}
To systematically investigate the occurrence and quality of Android related code snippets on \so{}, we downloaded\footnote{
archive.org offers the option to download an archive of all \so{} posts from their website, cf. \url{https://archive.org/details/stackexchange}} a dataset of all \so{} posts in March 2016, which gave us a dataset of 29,499,660 posts.
We extracted all posts which were tagged with the \texttt{android} tag - this resulted in 818,572 question threads with 1,165,350 answers. Questions in our data set had 1.4 answers on average. The oldest post in the dataset was from August, 2008. 559,933 (68.4\%) of the questions and 744,166 (63.9\%) of the answers contained at least one code snippet. Posts had 1,639.4 views on average. The most popular post in our dataset had 794,592 visitors.

With the oracle-based parser (as described in Section \ref{sec:parser}) we filtered the 818,572 questions and the 1,165,350 answer posts from \so{} which revealed 2,504 (2,474 distinct) security-related snippets from question posts and 1,517 (1,360 distinct) security-related code snippets from answer posts, respectively. In summary, using the JavaBaker oracle, we could successfully identify security-related snippets as shown in Table~\ref{table:identified-snippets}. 

The majority of snippets (2,841, i.e. 70.7\%) were related to the \textit{java.security} 
API which implements access control, generation/storage of public key pairs, message digest, signature and secure random number generation. Most snippets were related to cryptographic key initialization, storage (e.\,g. \textit{java.security.Key}, \textit{java.security.KeyPairGenerator} or \textit{java.security.KeyStore} -- 44.9\%) and message digests (\textit{java.security.MessageDigest} -- 30.4\%). This attunes to our intuition, as almost all cryptographic implementations require key management and hash functions are cryptographic primitives.

Code containing Android's cryptographic API was second most prevalent and present in 1,286 (31.9\%) code snippets. 1,088 (84.6\%) of these code snippets applied the \textit{javax.crypto.Cipher} API and hence, contained code for symmetric encryption/decryption. Interestingly, many snippets employ user-chosen raw keys for encryption (701 snippets with \textit{SecretKeySpec}) instead of generating secure random keys by using the API (207 snippets with \textit{KeyGenerator}). This indicates that most of the keys are hard-coded into the snippet, which states a high risk of key leakage if reused in an application due to reverse engineering.

The TLS/SSL package \textit{javax.net.ssl} was used in 28.9\% of the code snippets. The majority of these code snippets (545, i.e. 46.7\%),
contained custom TrustManagers to implement X.509 certificate validation. Optimistically, by implementing a custom trust manager, developers might aim at higher security by only trusting their own infrastructure. Practically, we observe that custom trust managers basically ignore authentication at all \cite{Fahl:2012uh}. 17.1\% of the code snippets contained custom hostname verifiers. Apache's SSL library was mainly used for enabling deprecated hostname verifiers that turned off effective hostname verification.

Code snippets containing code for BouncyCastle, SpongyCastle and SUN were rarely found. This could be due to the fact that those libraries are mostly called directly by only changing the security provider. Interestingly, nearly no (0.3\%) snippets contained code for the easy-to-use jasypt and keyzcar libraries. Possible reasons could be their low popularity or good usability. Similarly, the GNU cryptographic API was rarely used. This might be due to the difficulty to integrate it in an Android application~\cite{González2015}.

\begin{table}[htb!]
\centering
\resizebox{\linewidth}{!}{%
\begin{tabular}{p{0.01cm} l r p{0.01cm} l r} \hline \hline
\ml{2}{Namespace}                       &  Snippets & \ml{2}{Namespace}                       &  Snippets   \\ \hline \hline
\ml{2}{javax.crypto}                           &   1,286 & \ml{2}{android.security}                       &   5         \\
                        & Cipher                & 1,088  & \ml{2}{com.sun.security}                       &   5  \\
                        & KeyGenerator          & 207    & \ml{2}{gnu.crypto}                             &   47\\
                        & spec.SecretKeySpec    & 701    & \ml{2}{java.security}                          &   2,841\\
                        & spec.PBEKeySpec       & 69     & \ml{2}{javax.security}           &   44 \\
                        & spec.DESedeKeySpec    & 6      & \ml{2}{javax.xml.crypto}         &   3\\
                        & spec.DESKeySpec       & 21     & \ml{2}{org.bouncycastle}         &   48\\
                        & spec.IvParameterSpec  & 338    & \ml{2}{org.spongycastle}         &   44\\
                        & spec.RC2ParameterSpec & 1      & \ml{2}{org.jasypt}               &   11\\
                        & Mac                   & 85     & \ml{2}{org.apache.http.conn.ssl}                &  241\\
                        & Sealed                & 8      &  & AllowAllHostnameVerifier & 184\\
                        
\ml{2}{javax.net.ssl}                           &   1,166 && StrictHostnameVerifier & 51\\
                        & TrustManager          &   545  & & BrowserCompatHostnameVerifier & 8\\
                        & HostnameVerifier      &   200  && TrustSelfSignedStrategy & 1  \\
                        & SSLSocket             &   533  && SSLSocketFactory & 105  \\
\ml{2}{org.keyczar}              &   2&&&           \\
                          \hline \hline
\end{tabular}
}
\caption{Snippet counts per library.}
\label{table:identified-snippets}
\end{table}

\subsection{Evaluation of Code Classification}
\label{sec:coding-results}
Altogether, we classified 1,360 distinct security-related code snippets to provide a training set for the machine learning based classification model. This set contains all security-related code snippets we found in Android-related answer posts. We then applied the trained classifier on the complete set of 3,834 distinct security-related code snippets found in Android posts, including both questions (64.53\%) and answers (35.47\%). The security classification results of the training set are presented first and are described as follows:

The qualitative description of the snippets is divided into the security categories TLS/SSL, symmetric cryptography, asymmetric cryptography, random number generation, message digests, digital signatures, authentication, and storage.
For each category, we describe why we consider the respective code snippets to be insecure, what has been done wrong and why it (supposedly) has been done wrong. Whenever possible, we give counts for security mistakes and examples for the security mistakes we found.

Second, we demonstrate the feasibility of our state vector machine approach by discussing the overall quality of our classification model regarding precision, recall, and accuracy. Finally, we present the results for the large scale security classification of all security-related code snippets found on \so{}. \hfill\\

\subsubsection{Labeling of Training Set}
As described above, the training set consists of code snippets that have been identified by the oracle-based filter to include security-related properties (cf. Section \ref{sec:parser}). We classified a fraction of this set manually in order to provide supervision to the SVM. Subsequently, the SVM was able to classify the whole security-related set provided by the filter.\hfill\\

\paragraph{TLS/SSL}
We found 431 (31.48\%) of all snippets in the training set to be TLS related, among these we rated 277 (20.23\%) as insecure. In other words, almost one third of security-related discussions seem to target communication security and more than half of the related snippets would introduce a potential risk in real-world applications.
The majority of TLS snippets are insecure because of using a default hostname verifier or overriding the default TrustManager of \textit{java.net.ssl} when initializing custom TLS sockets. Every single custom TrustManager implementation we found consists of empty methods that disable certificate validation checks completely, while none of the custom TrustManager are used to implement custom certificate pinning, which is the reasonable and secure use case for creating custom TrustManagers. This correlates to our assumption stated in Section \ref{sec:eval_code_extract_filter}.
An empty TrustManager is implemented by 156 snippets, while 6 snippets use the \textit{AllowAllHostNameVerifier} - and 2 implemented both.
We found 42 snippets that override the verification method of \textit{HostnameVerifier} of \textit{java.net.ssl} by returning \texttt{true} unconditionally, which ultimately disables hostname verification completely (cf. Listing \ref{listing-example-verify}). This change to the HostnameVerifier implements the same behavior as \textit{AllowAllHostNameVerifier}.

We found several snippets that modify the list of supported ciphers. In all cases, insecure ciphers were added to the list. We assume this is caused by reasons of either legacy or compatibility.\hfill\\

\paragraph{Symmetric Cryptography}
We found 189 (13.80\%) of all snippets in the training set to be related to symmetric cryptography, among these we rated 159 (11.61\%) of the snippets as insecure.
For example, we found snippets containing encryption/decryption methods with less than 5 lines of code, implementing the minimum of code needed to accomplish an encryption operation. These snippets were insecure by using the cipher transformation string "AES" which uses ECB as default mode of operation (cf. Section~\ref{par:symcrypto}). Developers might be unaware of this default behavior or of ECB being insecure.

Another example are snippets that create raw keys and raw IVs using empty byte arrays (i.\,e. byte arrays which consist of zeros only), derive raw IVs directly from static strings, or by using the array indexes as actual field values as shown in Listing \ref{listing-example-empty-key}. Other snippets derive raw keys directly from strings that were mostly simple and insecure passphrases, e.g. \textit{"ThisIsSecretEncryptionKey"}, \textit{"MyDifficultPassw"}. We also found snippets that initialized the IV using the secret key. \hfill\\

\paragraph{Asymmetric Cryptography}
We found 59 (4.3\%) of all code snippets in the training set to include asymmetric cryptography API calls, among these 13 (0.94\%) of the snippets were classified as insecure. Considering the importance of public key cryptography in key distribution and establishing secure communication channels, 4.3\% is quite low and corresponds to our assumption in Section \ref{sec:eval_code_extract_filter}. All insecure snippets used weak key lengths which varied between 256 and 1024 bits for RSA keys. Obviously, recommendations from public authorities (e.g. the NIST) regarding secure cryptographic parameters are not fully taken into consideration. \hfill\\

\paragraph{(Secure) Random Number Generation}
We found 30 (2.19\%) of all code snippets in the training set to include (secure) random number generation API calls, among these 29 (2.11\%) of the snippets were classified as insecure. All insecure snippets explicitly seeded the random number generator with static strings (cf. Listing~\ref{listing-example-string-seed}). Replacing the random number generator's seed this way and not supplementing it results in low entropy~\cite{java-seeding}. \hfill\\

\paragraph{Digital Signatures and Message Digests}
Overall, 279 (20.37\%) snippets contain digital signatures related API calls. We classified none of them as insecure. This is a remarkable and unexpected observation, especially compared to the high number of insecure snippets in discussions regarding symmetric cryptography. To explain this we had a closer look at the relevant code snippets: calls to the digital signatures API are most often related to extracting existing signatures, not to validate them or generate new ones. Such an interactive query of existing signatures is not very error-prone regarding security. 

Further, we found 392 (28.63\%) snippets to contain message digest related API calls, among these 14 (1.02\%) were classified as insecure due to usage of weak hash algorithms. Again, compared to the quantity of insecure snippets of other categories this is quite a low percentage. In generating a message digest the biggest pitfall is choosing a weak hash function. We assume that state-of-the-art hash functions are relatively established in the \so{} community. \hfill\\

\paragraph{Remaining}
19 (1.38\%) of the snippets contained authentication code, where one snippet was classified as insecure. Eight (0.58\%) contained secure storage code, where three snippets were classified as insecure.\hfill\\

\paragraph{Not Security-Relevant}
We classified 342 snippets as not security-relevant as defined in~\ref{sec:rationale-oracle}.\hfill\\

\subsubsection{Model Evaluation of SVM Code Classifier}
We report our model evaluation on binary SVM classifier trained on the labeled training data set. Overall, after removing some duplicates, the training data set consists of 1360 samples, out of which 420 code snippets are identified as insecure. As introduced in Section \ref{sec:tfidf}, we use a \textit{tf-idf} vectorizer to convert code snippets into numeric vectors for training.

\begin{figure}[h!]
    \centering
    \includegraphics[width=0.49\textwidth]{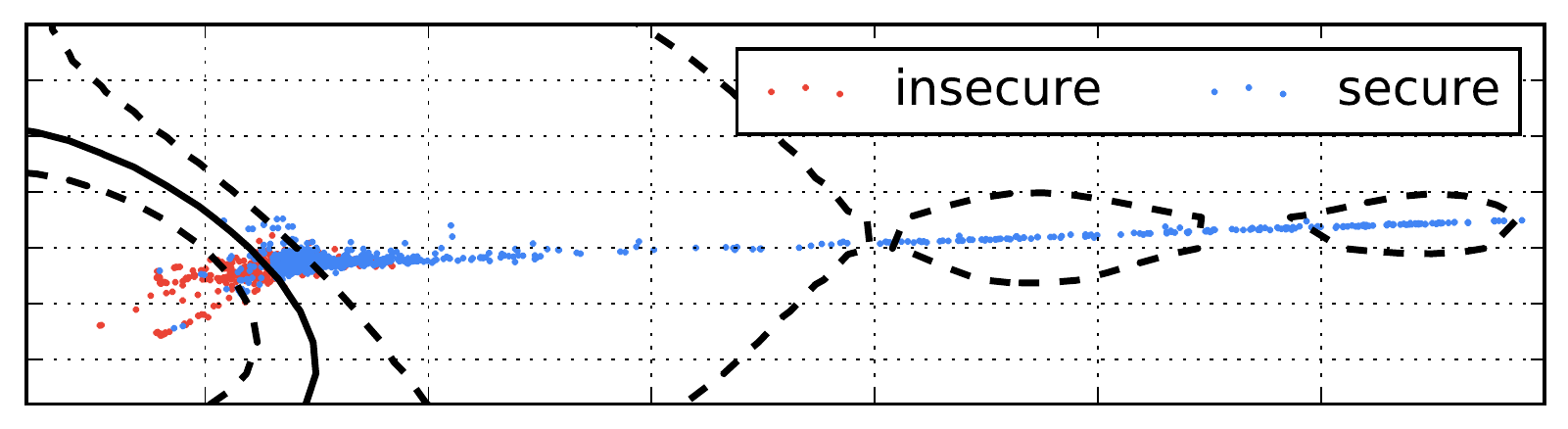}
    \caption{SVM with RBF kernel is trained on the training dataset, where the high dimensional training samples are projected on 2-dimensional using PCA. Solid contour line represents the classification boundary and dashed lines indicate the maximal margin learned by SVM. Insecure code snippets are marked as red circles, and secure ones are marked as blue circles.}
    \label{fig:svm_on_2dsamples}
\end{figure}

To better illustrate how SVM works, we show in Figure \ref{fig:svm_on_2dsamples} a demonstration in 2$d$ dimension, where we project the training samples in 2$d$ space by a common dimensionality reduction method, \textit{i.e.,} Principle Component Analysis (PCA) \cite{Smith2002}. The \textit{PCA} algorithm preserves two of the most informative dimensions by transforming the original coordinate system, and then a binary SVM is trained on the transformed data set. 
We leverage a RBF (Radial Basis Function) 
kernel function to tackle the non-linearity hidden in the projected training samples. The RBF kernel is a well known type of kernel to model non-linearity of data. It maps the non-linear input data to a high dimensional linear feature space, such that the data becomes linearly separable. We can see that even in 2$d$ space where some relevant information might be lost,  the SVM classifier can still produce a relatively good class boundary for both the secure (blue dots) and insecure (red dots) code samples.

Next, we evaluate our SVM model quantitatively by cross validating the training data set. First, we conduct a grid search on SVM to estimate the optimal penalty term $C$ (cf. (\ref{eq:svm_obj})) with respect to classification accuracy. Since the training data contains very high dimensional features, we use a linear kernel for SVM instead of RBF kernel in previous 2$d$ demonstration. The optimal parameter $C$ is determined by cross validation to be $0.644$ that is then fixed for the next evaluation steps. 
We evaluate the model on various training sizes with respect to precision, recall and accuracy. A discussion on these evaluation metrics can be found in \cite{Makhoul99performancemeasures}. In our experiment setup, we consider insecure samples as positive and secure ones as negative. Therefore, the precision score measures how many predicted insecure snippets are indeed insecure, recall score evaluates how many real insecure snippets are retrieved from all insecure snippets, and finally the accuracy score measures an overall classification performance taking both positive and negative samples into account.  

\begin{figure}[h!]
    \centering
    \includegraphics[width=0.45\textwidth]{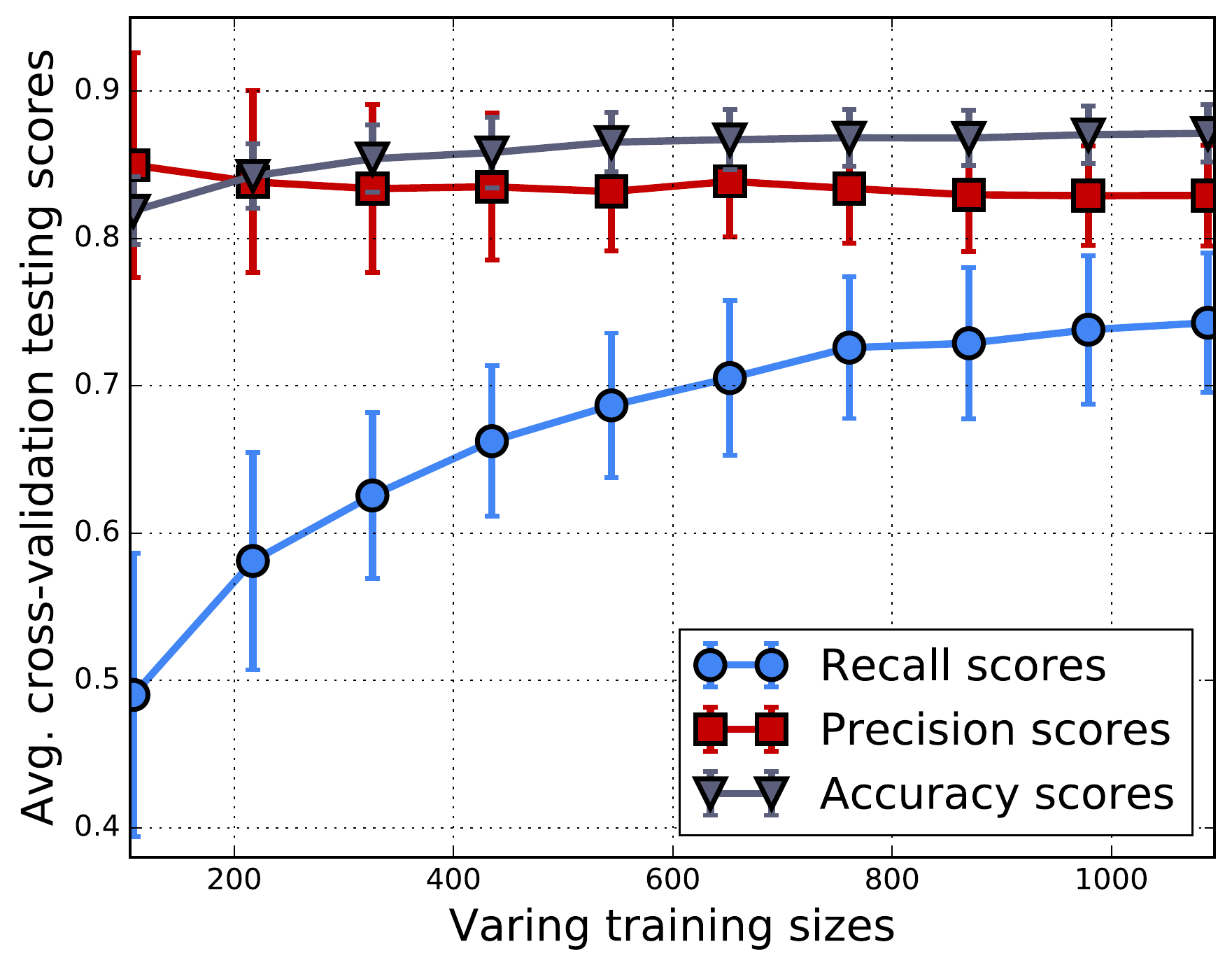}
    \caption{Binary SVM with linear kernel is trained over varying training sizes. 
     Cross validation is performed on each of these subsets of the training data set and evaluated with respect to precision, recall and accuracy scores.}
    \label{fig:svm_learn_curve}
\end{figure}

In Figure \ref{fig:svm_learn_curve}, we report the learning curve of our model with respect to varying training sizes. For each training size, a subset of the training data set is classified  
by the model with a 50-repetition cross validation. In each repetition, we randomly hold out 20\% of the training samples as testing set, and train on the remaining samples. Then, we evaluate the metrics respectively on the test set. Finally, we average the testing scores on all repetitions and plot the mean scores with 
standard deviation as the error bar. The results present a good precision and accuracy on varying training sizes, as the mean scores are approximately all above $0.8$. The constantly developing precision curve shows us that our model performs very well on detecting real insecure snippets instead of introducing too many false positives, even on a small training size. On the other hand, we see the recall curve is relatively poor on small training size. However, it reaches nearly $0.75$ when we have more than $1000$ training samples. Accuracy also improves with increasing training samples. The variance of the accuracy is canceled by combining both precision and recall. 

For completeness, we conduct a 5-fold cross validation on the whole training data set with optimal penalty term $C=0.644$. We report the confusion matrix of the best fold in Table \ref{table:conf-mat}. Note that the test size for each fold is 272. 

\begin{table}[h!]
\centering
\begin{tabular}{lcc} \hline \hline
True/Predicted          &   Secure (-1)   &   Insecure (+1)                    \\ \hline \hline
Secure (-1)                  &              181                  &       7       \\
Insecure (+1)                &  19                  &   65    \\ \hline
Summary &   accuracy: 0.904  &    precision: 0.903  \\ \hline
\end{tabular}
\caption{Confusion matrix}
\label{table:conf-mat}
\end{table}

To conclude our model evaluation, we argue that our SVM model can be even improved by conducting more effort on feature engineering and also by increasing the size of the training data set. In our experiment, we only remove comments in code snippets as a preprocessing step. In practice, this will be enhanced by applying a more complex token parser, e.g., static code parser, to generate better quality of features. Moreover, we did not leverage control flow information, which is considered informative of predicting security level, to enhance the model. 
A trivial refinement could be that we encode the relative position of each token in the snippet into the features. However, this could double the size of input feature dimension. Due to the complexity of model pruning and limit of training sample size, we decided to leave it for future work. Given the fact that the performance of the SVM model already achieves a level of practicability, we think that machine learning based methods can be a very strong supplementary for security code analysis.

\subsubsection{Large Scale Classification}
We applied the SVM code classifier on the complete set of 3,834 distinct security-related snippets from \so{}. Overall, we found 1,161 (30.28\%) insecure snippets and 2,673 (69.72\%) secure snippets. Out of the 1,360 distinct snippets found in answer posts, 420 (30.88\%) snippets were classified as insecure and and 940 (69.12\%) as secure. For the 2,474 distinct snippets we detected in questions posts, 741 (29.95\%) snippets were classified as insecure and 1,733 (70.05\%) as secure. 

\subsection{Evaluation of Code Detection}
\label{sec:largescale}
We applied our toolchain for code snippet detection (cf. Section \ref{sec:finding-code-snippets-in-apps}) on a large corpus of free Android applications from \googleplay{}. Beginning in October 2015, we successfully downloaded 1,305,820 free \androidapps from \googleplay{}\footnote{cf. \url{https://play.google.com}}. We re-downloaded new versions of apps until May 2016. The majority of apps received their newest update within the last 12 months.\hfill\\

\subsubsection{Apps with copied and pasted code snippets}
\label{sec:apps-with-cp-code}
Overall, we detected copied and pasted snippets in 200,672 (15.4\%) apps. Of these apps, 198,347 (15.2\%) contain a question snippet and 40,786 (3.1\%) apps contain an answer snippet. An overwhelming amount of apps contain an insecure code snippet: 196,403  (15\%) apps contain at least one. The top offending snippet has been found in 180,388 (13.81\%) apps and is presented in Listing~\ref{listing-empty-tm}. The remaining insecure snippets were found in 43,941 (3,37\%) distinct apps.

We found 506,922 (38.82\%) apps that contain a secure snippet. The most frequent secure snippet was detected in 408,011 (31.24\%) apps while the remaining snippets were contained in less then 73,839 (5.65\%) apps.
On average, an insecure snippet is found in 4,539.96 apps, while a secure code snippet is found in 10,719.83 apps. 

To investigate insecure snippets that were detected by our fully automated processing pipeline in detail, we performed a manual post-analysis of the categories described in Section \ref{sec:scoring}. To be more precise, we examined all security-related snippets that were detected in applications and sorted them by category. In the following, we give counts for affected applications for each security category. The given percentage values are related to applications that contain a snippet from \so{}. Further, we discuss the most offending snippets and estimate their practical exploitability.\hfill\\

\subsubsection{SSL/TLS}
The highest number of apps that implemented an insecure code snippet used this snippet to handle TLS. 183,268 (14.03\%) apps were affected by insecure TLS handling through a copied and pasted insecure code snippet. 
Conversely, only 441 (0.03\%) of all apps contained a secure code snippet related to TLS. 
For the large majority of 182,659 (13.98\%) apps with an insecure TLS snippet, their code snippet matches a question code snippet on \so{}, while only 22,040 (1.68\%) apps contain an insecure TLS snippet that was present in an answer on \so{}. A high risk example in this category is given by the top offending snippet as presented in Listing \ref{listing-empty-tm}, which uses an insecure custom TrustManager as described in Section \ref{sec:eval_code_extract_filter}. Missing server verification enables Man-In-The-Middle attacks by presenting malicious certificates during the TLS handshake. This snippet is a real threat with high risk of exploitation in the wild, as shown in \cite{Fahl:2012uh}.\hfill\\

\subsubsection{Symmetric Cryptography}
The second highest number of insecure code snippets in the wild were used for symmetric cryptography in 21,239 (1.62\%) apps. 
19,452 (1.48\%) of the apps with a code snippet that was related to symmetric cryptography had integrated a secure snippet. 
With a count of 19,189 apps, slightly more apps contain an insecure question snippet than an insecure answer snippet, which happened in 15,125 apps. The insecure snippet with the highest \cp{} count (found in 18,000 apps) within this category proposes AES in ECB mode. According to \cite{Egele:2013:ESC:2508859.2516693} this is vulnerable to chosen-plaintext attacks. Further, applications that include snippets with hard-coded cryptographic keys can most often be reverse-engineered without much effort. This leads leads to key leakage and therefore states a high risk (at least in the case where the key is not obfuscated).\hfill\\

\subsubsection{Asymmetric Cryptography}
We found only 114 (0.01\%) apps that contained an insecure code snippet related to asymmetric cryptography
, 698 (0.05\%) apps contained a secure asymmetric cryptography related snippet.
114 apps with insecure snippets contain an insecure question snippet.
29 apps implemented a secure answer snippet, 688 a secure question snippet.\hfill\\

\subsubsection{Secure Random Number Generation}
8,228 (0.63\%) apps contain an insecure code snippet related to random number generation, 
while 4,100 (0.31\%) apps contained a secure snippet. 
Most insecurities in this category come from question snippets (this was true for 8,227 apps, 
while 7,991 apps contain an insecure answer snippet).\hfill\\

\subsubsection{Hashes}
For hash functions, the majority of apps containing code snippets from \so{} contained secure code snippets: This was true in 4,012 (0.3\%)
, 14 apps contained an insecure one.\hfill\\

\subsubsection{Signatures}
15 apps contained a secure signature related snippet, while no insecure snippet was found in apps in this category. All of those snippets 
could be found in questions on \so{}.\hfill\\

\subsubsection{Not Security-Related}
\label{sec:not-sec-related}
Some of the snippets that were detected in apps could not be assigned to one of the categories above because they were not security-related as described in \ref{sec:rationale-oracle}.
498,046 (38.1\%) apps contained a snippet that was not security-related and therefore classified as secure. 

The most frequent secure snippet found in 408,011 apps was also not security-related. Therefore, considering security-related snippets only, we can state that significantly more Android applications contain an insecure snippet (196,403) than a secure one (73,839) (cf. Section \ref{sec:apps-with-cp-code}). \hfill\\

\subsubsection{Sensitive App Categories}
The largest number of sensitive apps that use insecure copied and pasted code snippets are 14,944 business apps, 4,707 shopping apps, and 4,243 finance apps. 
We find this result rather surprising, as we would have expected that security receives special consideration for these types of applications. Especially, finance apps have access to bank account information and therefore we would have expected them to be developed with extra care. Security and privacy is especially important in apps that handle medical data, as leaked sensitive data can have a severe impact on users. We found 2,000 medical and 4,222 health\&fitness apps that copied and pasted vulnerable \so{} code. 
Apps that are used for communication (3,745 apps) and social media (4,459 apps) are also widely affected.

\begin{figure}[ht]
	\centering
    \includegraphics[width=0.48\textwidth]{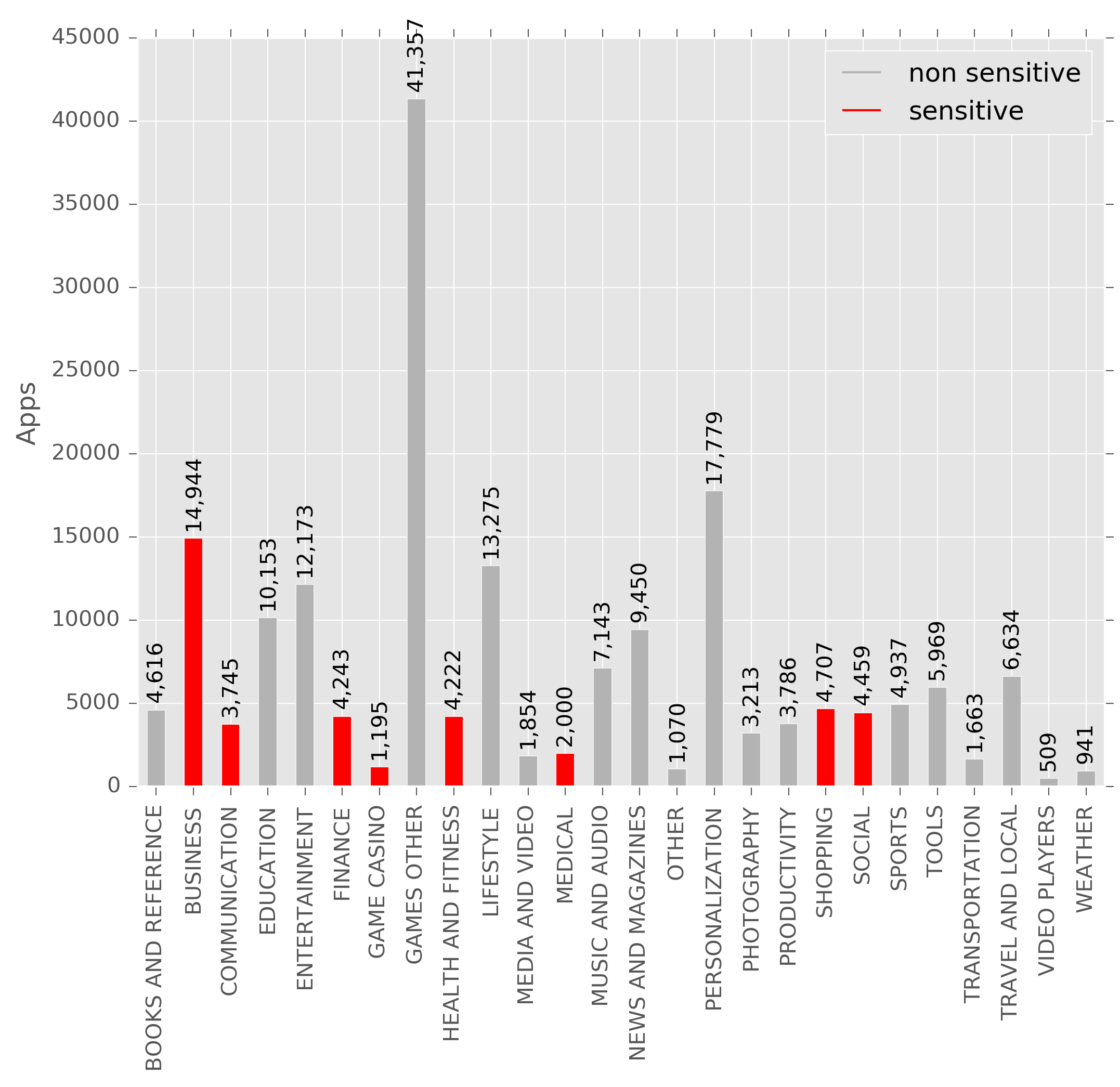}
	\caption{Distributions of insecure snippets found in Apps}
	\label{figDistribution}
\end{figure}

\subsubsection{Download Counts}
\begin{figure}[ht]
	\centering
    \includegraphics[width=0.48\textwidth]{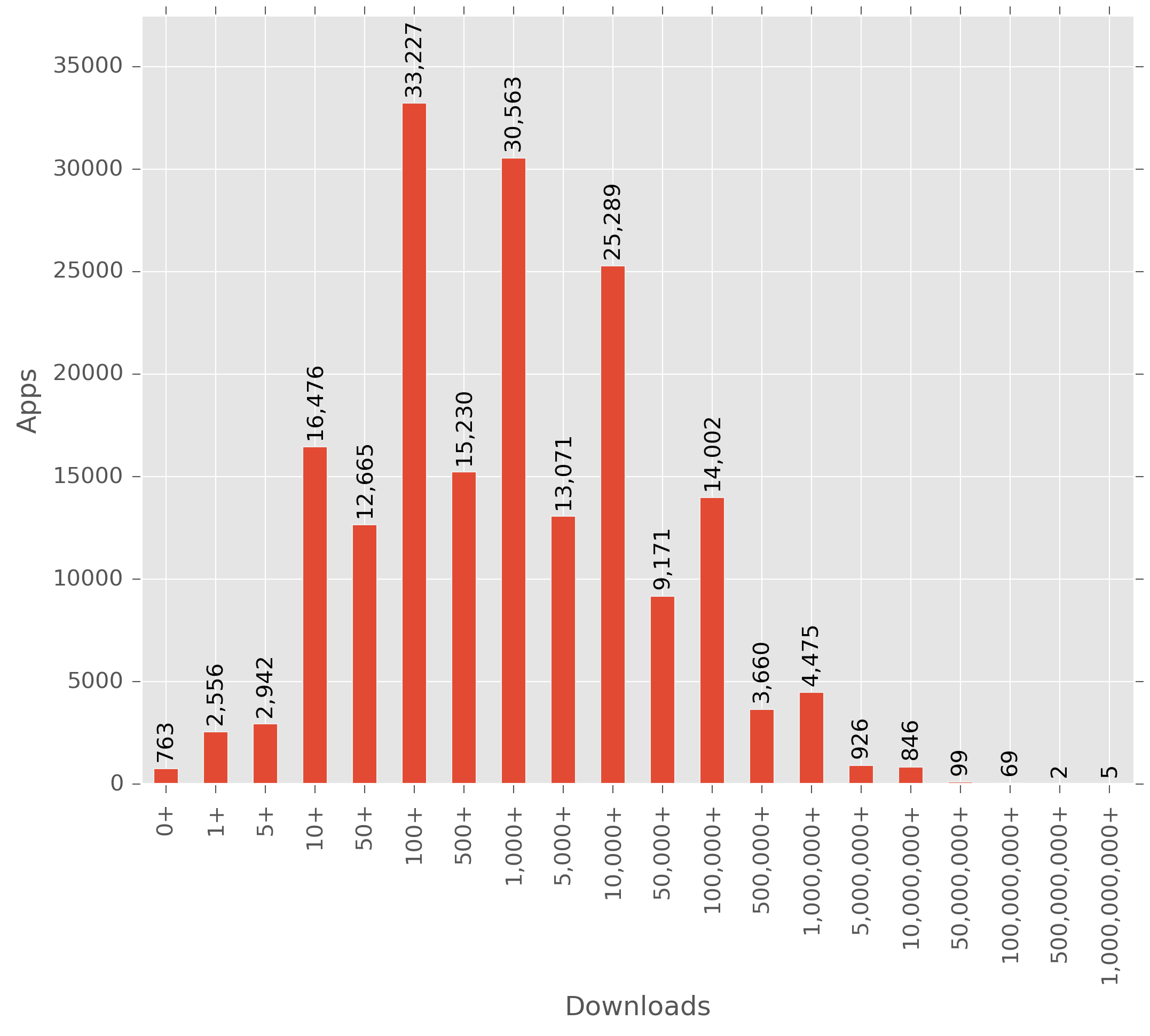}
	\caption{Download counts for Apps with insecure snippets}
	\label{figDownloads}
\end{figure}

In order to prove that we did not only inspect the long tail of unpopular apps provided by \googleplay{} we provide download counts for apps that contain insecure snippets in Figure~\ref{figDownloads}.

\subsection{Evaluation of Community Feedback}
\label{sec:evaluation:com_feedback}
For those insecure snippets that were detected in \androidapps{}, we analyzed the community feedback on \so{} as this represents the current public evaluation of posted code snippets. The available feedback mechanisms allow a very general evaluation of code snippets, e.\,g. with the presentation of view counters and the individual post score which results from up-/down-votes by the community. In addition to this, code snippets can be commented which is used to provide a more detailed feedback.

We analyzed if the existing feedback system provided by SO is capable of
informing the user about insecure snippets in an adequate way. At that, we analyze if the currently given feedback by the community is preventing or contributing to \cp{} of insecure code into \androidapps{}.\hfill\\

\subsubsection{Scoring}  
According to \so{}, a question snippet is supposed to be up-voted if it shows reasonable research effort in order to motivate the community to reply to it. Therefore, with pure focus on security aspects we expect insecure question snippets to be up-voted, as insecure code snippets intuitively demand more community research than secure ones. In contrast to question scores, answers are up-voted (according to \so{}) if they are estimated useful. Regarding the score of insecure answer snippets we expect a lower score as these do not provide a useful answer considering security-related snippets. 

The results in table~\ref{table:community-feedback} show that the scoring of insecure question snippets contradict to our assumption, because the secure ones have a higher score. In other words, the community assigns a higher needed research effort to questions with secure snippets. This is counter-intuitive from the security perspective and therefore leads to the conclusion that question scoring is not an adequate way of evaluating security. Of course, the community estimates needed research effort on the basis of a diversity of aspects, which outweigh security considerations. However, regarding answers, the low scoring of insecure snippets (as depicted in Table \ref{table:community-feedback}) correspond to our expectations. Again, this positive correlation might be caused by a variety of aspects, but from the security point of view it reveals the desired community behavior. However, aspects that are currently taken account for answer scoring are likely to be weighted differently in the future.

Next, we additionally include security warnings in our evaluation. Here, the scoring for questions given by the community are consistent with our intuition: Insecure questions including a warning are assigned with a higher scoring (corresponding to higher estimated research effort) than questions without such a warning. However, the scoring estimation regarding answer posts contradicts the desired community behavior: Insecure answers with security warning are scored significantly higher compared to the ones without warnings (cf. Table \ref{table:community-feedback}). Therefore, the influence of warnings (in answers) on the community score is highly questionable. A high scored answer with assigned security warning might confuse the reader.

The following considerations try an explanation of this result. Though a security warning should have a strong impact on the evaluation, the author of the warning can down-vote the score only once. On the one hand, this gives the community the ability to review the warning and to further reduce the score or to comment disagreement. On the other hand, the results show that the scoring of insecure snippets is partly contradicting and we did not find a single warning that has been questioned in a subsequent comment. Acar et al.~\cite{Acar:dev:2016} have shown that developers prefer functional snippets over secure snippets when implementing security related tasks in Android. This preference might also influence the scoring of security-related snippets which can result into a score that mostly considers functionality as the definition of a useful answer. 

When solely taking security considerations into account, we conclude that the currently deployed feedback system is insufficient for providing reliable and precise security estimation to the user.

\begin{table}[h!]
\centering
\resizebox{\linewidth}{!}{%
\begin{tabular}{lcccc} \hline \hline
Metadata            &   Secure   &   Insecure   &   Insecure+Warning  &   Insecure-Warning      \\ \hline \hline
Avg. Score Q/A      &  3.4/4.8   &    1.7/4.4   &   2.4/15.5          &   2.3/6.2                 \\
Avg. Viewcount Q/A  &  1467/4341 &    2254/8117 &   4081/16534        &   2812/10001              \\ \hline
\end{tabular}
}
\caption{Community feedback for security-related snippets regarding questions (Q) and answers (A)}
\label{table:community-feedback}
\end{table}

\subsubsection{Impact on \cp{}}
\label{sec:impact_copy_paste}

Next, we investigated if view count, warnings, and score of insecure snippets have an impact on the extent they are copied into applications. We first ordered all snippets according to their detection rate (the amount of applications that contained that snippet). For the snippets that ranked highest and lowest on this list (respectively 25\% -- we refer to these as top and bottom tier) we extracted the corresponding metadata from \so{}. This allowed us to observe possible correlations between view counts, warnings, and scoring to the actual \cp{} rate.

For both score and view count we found a positive correlation: A higher score or view count corresponds to an increased \cp{} count, as depicted in Table \ref{table:correlation-metadata-appcount}. This yields for both, questions and answers.

Interestingly, we see the opposite behavior with regard to warnings: Snippets that have been commented with security warnings are copied more often into applications than those without. An exceptionally striking example for this observation is the top offending snippet which was copied 180,388 times despite of being commented with warnings (cf. Listing \ref{listing-empty-tm}).

\begin{table}[h!]
\centering
\resizebox{\linewidth}{!}{%
\begin{tabular}{lcc} \hline \hline
Metadata                            &   Questions   &   Answers                 \\ \hline \hline
Avg. Score (top/bottom tier)        &   1.87/1.27   &   7.21/6.37               \\
Avg. Viewcount (top/bottom tier)    &   2,792/1,373 &   11,915/7805             \\\hline
\end{tabular}
}
\caption{Correlation of community feedback with \cp{} count of insecure code snippets}
\label{table:correlation-metadata-appcount}
\end{table}

\subsection{Limitations}
\label{sec:limitations_2}
Besides the limitations of the intermediate steps discussed in Sections \ref{sec:rationale-oracle} and \ref{sec:limit_2}, the overall processing pipeline does not fully prove that copied snippets originate from \so{}. To illustrate this objection, there theoretically could exist a third platform where snippets are copied and inserted to both, \so{} and \googleplay{}. However, \so{} is the most popular platform for developer discussions\footnote{\url{http://www.alexa.com/topsites}}. Further, \cite{Acar:dev:2016} et al. showed that developers most often rely on \so{} when solving security-related programming problems at hand. And finally, we found a positive correlation of the snippets view counts with their detected presence in applications (as discussed in Section \ref{sec:impact_copy_paste}). Therefore, it is very likely that snippets originate from \so{}.

\section{Related Work}
We focus on related work in four key areas, i.e. security of mobile apps, developer studies, investigation of \so{}, and detection of code reusage in apps.

\subsection{Security of Mobile Apps}
Fahl et al. analyzed the security of TLS code in 13,500 popular, free \androidapps{}~\cite{Fahl:2012uh}. They found that 8\% were vulnerable to Man-In-The-Middle attacks. In follow-up work, they extended their investigation to iOS and found similar results: 20\% of the analyzed apps were vulnerable to Man-In-The-Middle attacks~\cite{fahlccs:2013}. Oltrogge et al.~\cite{Oltrogge:usenix15} investigated the applicability of public key pinning in \androidapps{} and came to the conclusion that pinning was not as widely applicable as commonly believed. Egele et al.~\cite{Egele:2013:ESC:2508859.2516693} investigated the secure use of cryptography APIs in \androidapps{} and found more than 10,000 apps misusing cryptographic primitives in insecure ways. Enck et al.~\cite{Enck:2010:TIT} presented TaintDroid, a tool that applies dynamic taint tracking to reveal how \androidapps{} actually use permission-protected data. They found a number of questionable privacy practices in apps and suggested modifications of the Android permission model and access control mechanism for inter-component communication. Chin et al~\cite{Chin:2011} characterized errors in inter-application communications ({\it intents}) that can lead to interception of private data, service hijacking, and control-flow attacks. Enck et al.~\cite{Enck:2011ut} analyzed 1,100 \androidapps{} and reported widespread security problems, including the use of fine-grained location information in potentially unexpected ways, using device IDs for fingerprinting and tracking and transmitting device and location in plaintext. Poeplau et al.~\cite{kruegel:ndss14:dynloading} reported that many apps load application code via insecure channels allowing attackers to inject malicious code into benign apps.

\subsection{Developer Studies}
The bulk of identified security issues are attributed to developers that are poorly skilled in security-related programming. Core reasons for these issues were identified in a developer study
conducted by Fahl et al.~\cite{fahlccs:2013}: developers that customized TLS code disabled TLS functionality during testing and forgot to re-enable it for production, and they did not understand the security guarantees provided by and the security consequences imposed by improper TLS use. Similar root causes were reported by Georgiev et al.~\cite{Georgiev:2012ht}, showing that developers were confused by the many parameters, options and defaults of TLS APIs. Both papers explicitly mentioned \so{} as a platform that provides various solutions for "circumventing" TLS-related error messages by disabling TLS features. Acar et al.~\cite{Acar:dev:2016} conducted a laboratory study to investigate the impact of information sources on code security and found that developers using \so{} for looking up security-related issues produced the most functional but also the most insecure code, whereas participants using Android's official documentation produced more secure but less functional code.

\subsection{Investigation of \so{}}
Treude et al.~\cite{treude2011programmers} report that developer discussion platforms like \so{} are very effective at code reviews and conceptual questions. Vasilescu et al.~\cite{vasilescu2013stackoverflow} investigate the interplay of \so{} activity and development process on GitHub. They conclude that knowledge of the GitHub community flows into \so{}. In turn, this knowledge increases the number of commits of \so{} users on GitHub. Vasquez et al.~\cite{linares2014api} created an algorithm to link \so{} questions with Android classes detected in source code. They found that Android developer question counts peak on \so{} immediately after APIs receive updates that modify their behavior.

\subsection{Detection of Code Reusage in Apps}
Jiang et al.~\cite{Jiang:2007:DSA:1248820.1248843} compared the similarity of abstract syntax trees to detect code duplicates in source code. Hanna et al.~\cite{Hanna:2012:JSS:2481803.2481809} created k-gram streams from bytecode basic blocks. Each k-gram defines a program feature. A code snippet and an application is represented by the binary feature vector that is created using universal hashing over k-grams. They decide if a code snippet is contained in an app by dividing the number of common features by the number of features of the code snippet. While their approach works in benign scenarios, it is  not robust against trivial code modifications (e.\,g. reordering of instructions or renaming of variables). Crussell et al.~\cite{Crussell2012, Crussell2013} detect code clones by searching for subgraph isomophisms of program dependency graphs (PDG). Their approach is able to detect code fragments that perform similar computations through different syntactic variants~\cite{Chen:2014:AAS:2568225.2568286} and robust against trivial modifications, constant renaming and method/class restructuring. Chen et al.~\cite{Chen:2014:AAS:2568225.2568286} use Control Flow Graphs (CFG) in combination with opcodes to detect code clones in \androidapps{}. They define a geometry characteristic called centroid to embed a CFG into vector space.

\section{Countermeasures}
Now that we evaluated the extent of code flow from \so{} to \googleplay{} we discuss possible improvements for the current situation. On the one hand, there is a significant amount of secure code on \so{} that finds its way into real world applications. How can we reinforce this flow that surely is beneficial for the Android ecosystem? On the other hand, we also observed a vast amount of highly insecure code copied into critical applications. How can we prevent insecure code snippets from being copied?

In Section \ref{sec:evaluation:com_feedback} we showed that the deployed scoring system of \so{} is not fine-grained enough to mirror security concerns provided by the community. This suggests a scoring system that is purely focused on security aspects. However, a fine-grained scoring system will possibly also include equitable aspects such as code stability, efficiency, or audibility. This might impact the overall usability of \so{} and we fully understand the decision for just one score for each post.

Instead of expanding (and maybe complicating) the scoring system of posts, we propose another solution: Classification of code snippets into secure and insecure is fully automated in our approach, which allows us to implement a browser-plugin that directly indicates security issues by real-time classification of snippets. This includes both, snippets copied to the clipboard and snippets parsed on the actually watched discussion thread. Such a browser plugin is not limited to \so{}, but would work without much effort for any source of snippets in the web. We are currently actively developing such a browser plugin for Mozilla Firefox.

\section{Conclusion}

We present the first systematic and fully automated approach for measuring the flow of secure and insecure code from open developer platforms into Android market places. After scanning public discussion threads for code snippets and filtering the security-relevant fraction with a robust oracle-based filter, we apply machine learning classification to get a security scoring of relevant code snippets. By constructing an abstract representation in form of a program dependency graph for each snippet we detect their presence in closed source Android applications. Processing crowd-sourced code this way allows us to perform large-scale analysis of the proliferation of insecure code into large repositories.

We show the feasibility of our approach by scanning \googleplay{} for insecure code copied from \so{}. This choice is motivated by popularity and market dominance of both platforms, which serve as representative examples. We show that more than 196k of the 1.3 million applications from \googleplay{} contain vulnerable code copied from \so{} (cf. Section \ref{sec:limitations_2}). We detected 73k applications (cf. Section\ref{sec:not-sec-related}) using a secure code snippet from \so{}. By analyzing metadata we gain insight into developer behavior: From typical post up-voting trends and popularity of insecure code to favoured security libraries of specific domains (such as finance and gaming) we are able to draw interesting new conclusions on behavior of the Android developer community. We expect that a future systematic investigation augmenting metadata of security-related code snippets with metadata of their real-world clones in application repositories will serve as a rich source of new and interesting research questions.

So should \so{} be considered harmful? From classical risk evaluation perspective, the answer to this question depends on domain specific assets: A banking application with flawed cryptographic key initialization causes severe damage to the respective bank, even if the application has a relatively small user group. The same flaw in a set of popular gaming apps with very high download counts might not represent a major threat to the individual game developer studios, but has the potential to impact the Android ecosystem on a large scale. So depending on perspective, domain specific assets, and associated risks, the concrete threats posed by copying crowd-sourced code into applications must be evaluated individually. Finally, we want to stress the benefits of including secure code snippets into real-world applications. We identified several secure code snippets in critical applications, which is of great good for the community.

In a broad sense we infer conclusions regarding security of the Android ecosystem from comprehensive analysis of community discussions. We strongly believe that future research based on large scale data mining of community discussions will provide unexpected new and exciting insights into information security in general.

\section*{Acknowledgements}
The authors would like to thank Siddharth Subramanian for his strong support with JavaBaker and the anonymous reviewers for their helpful comments.

\bibliographystyle{IEEEtran}
\bibliography{refs}

\appendix

\lstset{
language=Java,
caption={Empty HostnameVerifier - Accepts all hostnames},
label=listing-example-verify,
basicstyle=\footnotesize\tt,
breaklines=true
}

\begin{minipage}[c]{0.95\linewidth}
\begin{lstlisting}[frame=single]
@Override
public boolean verify(String hostname, SSLSession session) {
    return true;
}
\end{lstlisting}
\end{minipage}

\lstset{
language=Java,
caption={Sample of static IVs and Keys in Snippets},
label=listing-example-empty-key,
basicstyle=\footnotesize\tt,
breaklines=true
}

\begin{minipage}[c]{0.95\linewidth}
\begin{lstlisting}[frame=single]
byte[] rawSecretKey = {0x00, 0x00, 0x00, 0x00, 0x00, 0x00, 0x00, 0x00, 0x00, 0x00, 0x00, 0x00, 0x00, 0x00, 0x00, 0x00};
String iv = "00000000";
byte[] iv = new byte[] { 0x0, 0x1, 0x2, 0x3, 0x4, 0x5, 0x6, 0x7, 0x8, 0x9, 0xA, 0xB, 0xC, 0xD, 0xE, 0xF };
\end{lstlisting}
\end{minipage}

\lstset{
language=Java,
caption={String used to replace the random number generators seed},
label=listing-example-string-seed,
basicstyle=\footnotesize\tt,
breaklines=true,
showstringspaces=false
}
\begin{minipage}[c]{0.95\linewidth}
\begin{lstlisting}[frame=single]
byte[] keyStart = "this is a key".getBytes();
SecureRandom sr = 
    SecureRandom.getInstance("SHA1PRNG");
sr.setSeed(keyStart);
\end{lstlisting}
\end{minipage}

\lstset{
language=Java,
caption={Top offending snippet},
label=listing-empty-tm,
basicstyle=\footnotesize\tt,
breaklines=true
}
\begin{minipage}[c]{0.95\linewidth}
\begin{lstlisting}[frame=single]
TrustManager tm = new X509TrustManager() {
    public void checkClientTrusted(X509Certificate[] chain, String authType) throws CertificateException { }
    public void checkServerTrusted(X509Certificate[] chain, String authType) throws CertificateException { }
    public X509Certificate[] getAcceptedIssuers() { return null; }
};
\end{lstlisting}
\end{minipage}

\end{document}